\documentclass[prl,aps,preprint,superscriptaddress]{revtex4}

\usepackage{graphicx}
\usepackage{float}
\usepackage[caption=false,subrefformat=parens,labelformat=parens]{subfig}

\begin{document}

\title[Strong ASOC and superconducting properties in LaPtSi]{Strong antisymmetric spin-orbit coupling and superconducting properties: The case of noncentrosymmetric LaPtSi}

\author{S. Palazzese}

\affiliation{Centro de F\'{\i}sica, Instituto Venezolano de
Investigaciones Cient\'{\i}ficas, Apartado 20632, Caracas
1020-A, Venezuela}

\affiliation{Departamento de F\'isica, Universidad Sim\'on Bol\'ivar, Apartado 89000, Caracas 1080-A, Venezuela}

\author{J. F. Landaeta}
\author{D. Subero}
\affiliation{Centro de F\'{\i}sica, Instituto Venezolano de
	Investigaciones Cient\'{\i}ficas, Apartado 20632, Caracas
	1020-A, Venezuela}

\author{E. Bauer}
\affiliation{Institut f\"{u}r Festk\"{o}rperphysik, Technische
Universit\"{a}t Wien, A-1040 Wien, Austria}

\author{I. Bonalde}
\affiliation{Centro de F\'{\i}sica, Instituto Venezolano de
	Investigaciones Cient\'{\i}ficas, Apartado 20632, Caracas
	1020-A, Venezuela}

\begin{abstract}
In this work we aim to analyze the effect of a strong antisymmetric spin-orbit coupling (ASOC) on superconductivity of noncentrosymmetric LaPtSi.  We study the energy gap structure of polycrystalline LaPtSi by using magnetic penetration depth measurements down to 0.02$T_c$. We observed a dirty s-wave behavior, which provides compelling evidence that the spin-singlet component of the mixed pairing state is highly dominant. This is consistent with previous results in the sense that the mere presence of a strong ASOC does not lead to unconventional behaviors. Our result also downplays LaPtSi as a good candidate for realizing time-reversal invariant topological superconductivity.
\end{abstract}




\maketitle


In noncentrosymmetric superconductors the absence of inversion symmetry in a crystal structure causes the appearance of an ASOC, which lifts the spin degeneracy that results in a possible admixture of spin-singlet and spin-triplet pairing states. ASOC thus leads to a spin-orbit band splitting $E_{SO}$ that is thought to have implications in the superconducting behavior, regardless of the electron pairing mechanism, when it is significantly larger than the superconducting energy scale $k_BT_c$ \cite{frigeri2004}. For instance, unconventional properties such as zeros in the superconducting gap function may appear in the presence of a strong ASOC. In this regard, line nodes in the superconducting gap structure of CePt$_3$Si \cite{bonalde2005,hayashi2006} and Li$_2$Pt$_3$B \cite{yuan2006}, both without inversion symmetry and with $E_{SO}/k_BT_c > 800$, were explained in terms of an admixture of spin states.

However, strong ASOC does not always imply unconventional behaviors. Noncentrosymmetric LaPt$_3$Si, with a larger $E_{SO}/k_BT_c$ than the one of isostructural CePt$_3$Si, has an isotropic energy gap structure consistent with a largely dominant $s$-wave pairing state \cite{ribeiro2009}. This is a remarkable difference with respect to CePt$_3$Si, whose line nodes may be related to an antiferromagnetic (AFM) order coexisting with superconductivity or to strong electronic correlations \cite{Fujimoto2006} (both conditions absent in LaPt$_3$Si). While in most noncentrosymmetric superconductors $E_{SO}>k_BT_c$, only in a few $E_{SO}/k_BT_c > 100$: LaPdSi$_3$ \cite{smidman2014}, BaPtSi$_3$ \cite{ribeiro2014}, PbTaSe$_2$ \cite{pang2016}, Th$_7$Fe$_3$ \cite{sereni1994}, CaIrSi$_3$ \cite{singh2014}, and K$_2$Cr$_3$As$_3$ \cite{pang2015}. With the exception of K$_2$Cr$_3$As$_3$, all these superconductors have a very strong ASOC, are nonmagnetic, have weak electronic correlations, and possess an isotropic superconducting energy gap consistent with a highly dominant spin-singlet pairing state. As in LaPt$_3$Si, unconventional behaviors are not seen in these materials. 

By now, there is mounting evidence that independently of the ASOC strength in noncentrosymmetric superconductors---as it happens in other superconducting families---the presence of magnetism seems to be essential for the emergence of unconventional behaviors (see discussion below) \cite{pang2015,Bauer2012,landaeta2017,landaeta2018}.

The appearance of nodes in the energy gap of noncentrosymmetric superconductors leads to another implication of a strong ASOC: candidacy for hosting time-reversal invariant nodal topological superconductivity \cite{Fujimoto2006,Sato2009,Schnyder2011,Schnyder2012,Schnyder2015}. The proper balance between the spin-singlet and spin-triplet components of the mixed pairing states may lead to topologically nontrivial line nodes, as has been suggested for CePt$_3$Si and Li$_2$Pt$_3$B among other noncentrosymmetric materials \cite{Schnyder2015}. In the nodal topological superconductors the properties of the topologically protected (Majorana) surface states are related to the bulk nodal structure and the symmetries of the order parameter \cite{Schnyder2015}. One should note that topological order can also appear in noncentrosymmetric superconductors with broken time-reversal symmetry (under the application of a magnetic field), in which case the topological phase is independent of the spin-triplet component of the mixed state \cite{Sato2009,Ghosh2010}.

Having a very strong ASOC ($130-255$ meV) \cite{kneidinger2013}, the weakly correlated nonmagnetic superconductor LaPtSi ($T_c=3.7$ K) without inversion symmetry is a good prospect to deepen further into the influence of an ASOC on the superconducting properties of noncentrosymmetric compounds. For this purpose it is quite relevant to study the gap structure and the symmetry of the order parameter. A previous work on specific heat in this compound suggested an exponential BCS-like behavior down to 0.67$T_c$ \cite{kneidinger2013}, indicative of a ruling spin-singlet component. However, due to the lack of information in the true low-temperature region (below 0.2$T_c$), the energy gap structure remains uncertain. 

Here, we present measurements of the magnetic penetration depth of LaPtSi down to 0.02$T_c$. It was observed a conventional $s$-wave behavior, which firmly states the dominance of the fully gapped spin-singlet component of the parity mixing in LaPtSi. In light of this result, we discuss further the role of the ASOC in the appearance of unconventional behaviors by comparing all previous relevant findings.


Polycrystalline LaPtSi was grown by arc melting the stoichiometric amounts of pure metal ingots under Ti-gettered argon and annealed under vacuum in sealed quartz ampoules at 800$^\circ$C for one week \cite{kneidinger2013}. An X-ray analysis ensures the sample purity since no other phases were observed \cite{kneidinger2013}. A sample was cutted to the dimensions $0.39\times0.25\times0.25$ mm$^3$ and then polished with aluminum oxide to minimize surface irregularities. It is worth noting that LaPtSi is highly fragile, which makes it difficult to obtain the appropriate dimensions without leaving out irregularities.

Penetration-depth measurements were carried out using a 13.5 MHz tunnel diode oscillator \cite{bonalde2005} down to around 50 mK. The magnitude of the magnetic ac field was estimated to be less than 5 mOe and the dc field at the sample was reduced to around 1 mOe. The deviation of the penetration depth from the lowest measured temperature, $\Delta\lambda(T)=\lambda(T)-\lambda(T_{min})$, was obtained up to $T \sim 0.99T_c$ from the change in the measured resonance frequency $\Delta f(T)$: $\Delta f(T) = G\Delta \lambda(T)$. Here $G$ is a constant estimated by measuring a sample of known behavior and of the same dimensions of the test sample.

\begin{figure}\begin{center}
	\scalebox{0.7}{\includegraphics{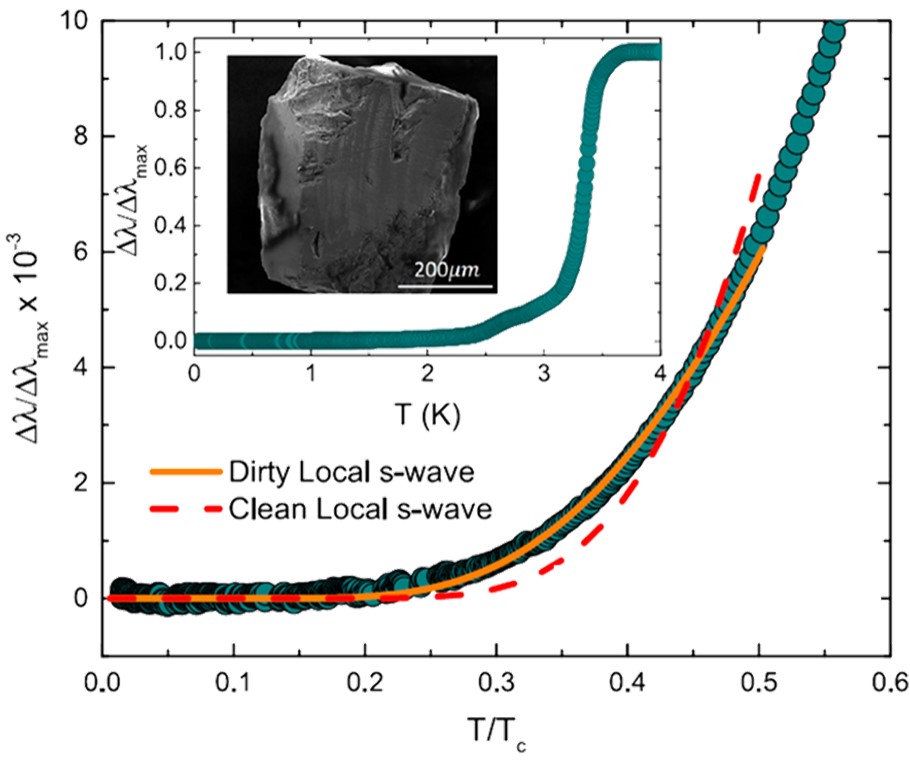}}
	\caption{\label{figure1}{Magnetic penetration depth of LaPtSi in the low-temperature regime fitted to clean and dirty local s-wave models. The inset shows the complete temperature behavior of the penetration depth and a photomicrograph of the sample.}}\end{center}
\end{figure}

The main panel of figure~\ref{figure1} shows the low-temperature variation of the magnetic penetration depth $\Delta \lambda(T)$ normalized to $\Delta \lambda_{max}$, defined as the total penetration-depth shift from $T_c$ down to the lowest temperature. The inset of figure~\ref{figure1} shows the entire superconducting region and a photomicrograph of the sample. The transition temperature $T_c$=3.74 K was taken at the onset of the diamagnetic transition, in good agreement with a previous report on resistivity and heat-capacity measurements \cite{kneidinger2013}. The small irregular behavior around 3 K may be due to surface irregularities (see micrograph in the inset of figure~\ref{figure1}). In figure~\ref{figure1} the error bars can be considered the size of the dots.

In figure~\ref{figure1} it is observed that the experimental data flatten out below 0.2$T_c$, as expected for a superconductor with an isotropic energy gap. LaPtSi is a dirty type-II superconductor \cite{ramakrishnan1995}, since the coherence length $\xi_{GL}(0)=338$ \r{A} is much larger than the mean free path $l=43$ \r{A}. We fitted the experimental data to clean and dirty s-wave models up to 0.5$T_c$. We used the low-temperature approximations of the clean local BCS model

\begin{equation}
\Delta\lambda(T)_{clean}\propto\sqrt{\frac{\pi\Delta_0}{2k_BT}}\exp(-\Delta_0/k_BT), 
\end{equation}

\noindent where $\Delta_0$ is the zero-temperature energy gap and $k_B$ is the Boltzmann constant, and the dirty local s-wave model \cite{tinkham}

\begin{equation}
\Delta\lambda(T)_{dirty} \propto\exp(-\Delta_0/k_BT).
\end{equation}

It is obvious from figure~\ref{figure1} that the dirty local s-wave model fits the data better than the clean model, reaffirming the dirty condition of this superconductor.  The dirty model yields a zero-temperature energy gap $\Delta_0=1.73k_BT_c$, which is very similar to the standard BCS value ($\Delta_0=1.76k_BT_c$). Thus, our results indicate that LaPtSi can be described in terms of s-wave superconductivity with an isotropic energy gap, confirming early heat-capacity measurements \cite{kneidinger2013}.

In regard to the possible pairing mixed states, our results indicate that the spin-singlet component is predominant in LaPtSi as well, in clear concordance with other noncentrosymmetric superconductors with very strong ASOC that show conventional behaviors.

\begin{table}
	\centering
	\caption{Some reference compounds with their critical temperatures, approximate maximum spin-orbit band splittings in meV, spin-orbit band splitting to superconducting energy scale ratios, and superconducting gap structures. Here F: fully gapped, L: line node, P: point node}
	\label{tabla}
	\begingroup
	\setlength{\tabcolsep}{15pt}
	\begin{tabular}{ccccc}
		\hline \hline
		Compound & T$_c$(K) & E$_{SO}$ (meV) &  E$_{SO}$/k$_B$T$_c$  & Gap structure\\ \hline
		LaPt$_3$Si & 0.64 & \textbf{69-207} \cite{Onuki2012} & 1250-3700 & F \cite{ribeiro2009}  \\ 
		CePt$_3$Si & 0.75 & 50-200 \cite{Onuki2012,samokhin2004} & 770-3100 & L \cite{bonalde2005}  \\ 
		BaPtSi$_3$ & 2.25 &  $\le$300 \cite{bauer2009} &  $\le$1500 & F\cite{ribeiro2014} \\ 
		PbTaSe$_2$ &3.72 & 114-444 \cite{ali2014,Bian2016} & 356-1400 & F \cite{pang2016}  \\
		LaPdSi$_3$ & 2.65 & $\le$200 \cite{winiarsky2015} & $\le$880 & F\cite{smidman2014}\\
		Li$_2$Pt$_3$B & 2.8 & $\le$200 \cite{lee2005} & $\le$830 & L \cite{yuan2006,takeya2005} \\  
		LaPtSi & 3.7  & 130-255 \cite{kneidinger2013}  & 410-800 & F $^{\rm a}$\\ 
		CaIrSi$_3$ & 3.5 & $\le$200 \cite{uzunok2016} & $\le$660 & F \cite{singh2014} \\
		Th$_7$Fe$_3$ & 2.09 & 40-100 \cite{sahakyan2017} & 220-555 & F \cite{sereni1994} \\
		K$_2$Cr$_3$As$_3$ & 6.1 & 60 \cite{jiang2015} & 114 & L \cite{pang2015}\\
		CeRhSi$_3$ & 1.05 $^{\rm b}$ & \textbf{12} \cite{terashima2008} & 133 & L \cite{landaeta2018} \\
		LaNiC$_2$ & 3.3 & \textbf{20} \cite{hirose2012} & 70 & P \cite{landaeta2017,Lee1996,Bonalde2011} \\
		CeIrSi$_3$ & 1.5 $^{\rm c}$ & \textbf{3} \cite{Onuki2012} & 23 & F,L \cite{landaeta2018,mukuda2008}\\
		Y$_2$C$_3$ & 18 & 2-15 \cite{yusuke2007} & 1-10 & F\cite{akutagawa2006}/L\cite{chen2011}\\ 
		\hline \hline
	\end{tabular}
	\endgroup
	\begin{flushleft}
		$^{\rm a}$ {\footnotesize Found in this work}
		\\$^{\rm b}$ {\footnotesize At 2.8 GPa}
		\\$^{\rm c}$ {\footnotesize At 2.6 GPa}
	\end{flushleft}
\end{table}

Table~\ref{tabla} shows a list of noncentrosymmetric compounds organized according to their $E_{SO}/k_BT_c$ strengths. Each material was selected following the criterion that either its ratio $E_{SO}/k_BT_c > 500$ or its energy gap has been found to possess nodes. It is notable that in general superconductors with an estimated strong ASOC display a conventional behavior unless magnetic order is present. Among the standard BCS superconductors are the nonmagnetic systems LaPt$_3$Si, BaPt$_3$Si, PbTaSe$_2$, LaPdSi$_3$, CaIrSi$_3$, Th$_7$Fe$_3$, and LaPtSi. Unconventional superconducting behaviors appear independently of the ASOC strength and in connection with a magnetic instability. Such is the case of CePt$_3$Si, K$_2$Cr$_3$As$_3$, CeRhSi$_3$, CeIrSi$_3$, and LaNiC$_2$ (recently found to have a magnetic phase at finite pressures \cite{landaeta2017}). 

Nonmagnetic Li$_2$Pt$_3$B seems to be at odds with this pattern, since line nodes in the energy gap have been found in magnetic penetration-depth \cite{yuan2006} and heat-capacity \cite{takeya2005,eguchi2013} measurements. Y$_2$C$_3$ may also be in conflict, although the two results reported so far in this compound are contrasting. Penetration depth indicates line nodes \cite{chen2011}, whereas heat capacity suggests an isotropic energy gap \cite{akutagawa2006}.

Thus far there is no plausible explanation for the appearance of unconventional behaviors as a consequence of magnetic instabilities. It may be that the presence of a finite magnetic moment somehow enhances the spin-triplet component of the parity mixing due to an ASOC \cite{Fujimoto2006,Yanase2007}.

With regard to topological order, our result of a fully gapped LaPtSi discards this compound as a candidate for a time-reversal symmetric topological superconductor. Notwithstanding, LaPtSi can still be considered a prospect for topological order under the application of an external magnetic field \cite{Sato2009,Ghosh2010}. It seems that good noncentrosymmetric candidates for nodal topological superconductivity would be those that have a nearby magnetic instability rather than a strong ASOC.

We performed magnetic penetration-depth measurements in noncentrosymmetric LaPtSi down to $0.02T_c$. We observed an s-wave BCS-like behavior in the local dirty limit. Our result reasserts that a strong ASOC does not seem to have the sole responsibility for unconventional behaviors in noncentrosymmetric materials. In these compounds as well, amply accepted results point to the need for proximity to a magnetic instability for the appearance of unconventional superconducting behaviors. The fully gapped state rules out LaPtSi as a good candidate for hosting time-reversal invariant nodal topological superconductivity, which may be more realizable in noncentrosymmetric compounds with some kind of magnetism.

\begin{acknowledgments}
	We thank M. Morgado from the Electron Microscopy Facility at the Venezuelan Institute for Scientific Research (IVIC) for support with the SEM and EDX studies. This work was supported by Austrian FWF (grant number P22295).
\end{acknowledgments}



\providecommand{\newblock}{}

\end{document}